\documentclass[12pt]{article}%
\usepackage{amssymb}
\usepackage{graphicx}
\usepackage{amsmath}
\usepackage{amsfonts}%
\newtheorem{theorem}{Theorem}

\begin{document}

\title{\textbf{MAGNETOSTRICTION TRANSITION}}
\author{Senya Shlosman\\Centre de Physique Th\'eorique, \\CNRS-Luminy-Case 907, \\F-13288 Marseille Cedex 9, France \\and \\IPPI, RAS, Moscow, Russia\\\textit{shlosman@cpt.univ-mrs.fr}
\and Valentin Zagrebnov\\Universit\'e de la M\'editerran\'ee \\and \\Centre de Physique Th\'eorique, \\CNRS-Luminy-Case 907, \\F-13288 Marseille Cedex 9, France \\\textit{zagrebnov@cpt.univ-mrs.fr}}
\maketitle

\begin{abstract}
We present a \textit{non mean-field} model which undergoes a magnetostriction
phase transition in the temperature. That is, the crystal becomes sharply
contracted and magnetized once the temperature passes below the critical value.

\end{abstract}

\section{Model and main theorem\label{section 1}}

Magnetostriction is know in physics as a phenomenon of a drastic change of
geometric shape of crystals, which is accompanied by magnetic transition, see
e.g. \cite{K}, \cite{M}. Usually it is a first order phase transition with a
jump of spontaneous magnetization together with the jump in geometry of the
crystal elementary cells. Physical origin of this phenomenon is related to
so-called magnetoelastic coupling, i.e. to the interaction between spin and
displacement degrees of freedom in magnetic crystals, \cite{K}. Various
mean-field theories of this phenomenon were discussed in literature since a
long time. See, e.g., \cite{ZF} and references therein, for crystals, and
\cite{GZ} for magnetosriction in ferrofluids. (The solvable model with a
short-range interaction, discussed in \cite{M}, does not exhibit the jump
specific for magnetostriction, because it is one-dimensional.)

In the present paper we propose a simple - and a first non mean-field! - model
of this phenomenon. We prove that our model undergoes the phase transition,
when the crystal becomes sharply contracted and magnetized, once the
temperature passes below the critical value, provided the dimension is at
least two.

We consider the following model: at each site $s$ of $\mathbb{Z}^{d}$ we have
an Ising spin \ $\sigma_{s},$ while at each bound $l=\langle st\rangle$ of the
lattice we have positive real variable $r_{st}$, playing the role of the
spatial distance between two sites.

Initially we were interested in the Hamiltonian
\[
\tilde{H}_{\Lambda}(\sigma^{\Lambda},r^{\Lambda_{b}})=-\sum_{\langle
st\rangle\in\Lambda_{b}}J(r_{st})\ \sigma_{s}\sigma_{t}+\mu\sum_{\langle
st\rangle\in\Lambda_{b}}(r_{st}-R)^{2}-h\sum_{s\in\Lambda}\sigma_{s}\text{.}%
\]
Here the function $J\left(  \cdot\right)  \geq0$ describes the dependence of
the strength of the interaction between the spins $\sigma_{s}$ and $\sigma
_{t}$ on their spatial separation. The parameter $R$ is the ground-state
distance between sites in the absence of the spin interaction; $h$ is the
external magnetic field. We were assuming that $J$ is small on large distances
and large on small distances. Our hope was to show that in the symmetric case
-- $h=0$ -- the model would undergo the striction transition as the
temperature goes down. But we were unable to show that, and, moreover, our
computations suggest that such first order transition does not take place for
the Hamiltonian $\tilde{H}.$

To realize our program we have to modify our Hamiltonian, adding another
``geometric'' term to the interaction. Namely, we will consider the model,
defined by the following Hamiltonian:
\begin{align}
H_{\Lambda}(\sigma^{\Lambda},r^{\Lambda_{b}})  &  =-\sum_{\langle st\rangle
\in\Lambda_{b}}J(r_{st})\ \sigma_{s}\sigma_{t}+\mu\sum_{\langle st\rangle
\in\Lambda_{b}}(r_{st}-R)^{2}\label{04}\\
&  +\lambda\sum_{\langle st\rangle,\langle s^{\prime}t\rangle\in\Lambda
_{b}:\left|  s-s^{\prime}\right|  =\sqrt{2}}(r_{st}-r_{s^{\prime}t})^{2}%
-h\sum_{s\in\Lambda}\sigma_{s}\text{.}\nonumber
\end{align}
Here in addition to the parameter $\mu>0,$ which is enforcing the lattice
structure with the spacing to be close to $R,$ we add another parameter
$\lambda>0,$ which has the effect of making the $r$-lattice more regular. In
particular, this term makes the ``triangle inequality violation''
energetically unfavourable. By the ``triangle inequality violation'' we mean,
for example, the situation when among the four bonds $r_{st},r_{s^{\prime}%
t},r_{s^{\prime}t^{\prime}},r_{st^{\prime}},$ forming a plaquette of the
lattice, there are three relatively small values and one relatively big.

To ensure that the above model undergoes the striction transition we have to
suppose that the interaction $J$ is weak enough on large distances $r,$ and is
strong enough on small distances. Otherwise this function can be fairly
general. We will describe now one specific choice of the class of interactions
$J,$ for which the transition takes place; other choices are also possible.

We are supposing that above some value $\rho>0$ the interaction is weak:
\[
J\left(  r\right)  \leq u\text{ for }r\geq\rho,
\]
with $u$ small. We further suppose that the interaction is bounded:
\[
\max_{r>0}J\left(  r\right)  =\bar{U}<\infty,
\]
and that within the region $r\leq\rho$ it is sufficiently strong: for some
$K\subset\left[  0,\rho\right]  $ and for all $r\in K$%
\[
J\left(  r\right)  \geq U,
\]
with $U$ large, while $\frac{\bar{U}}{U}=1+\varkappa$ with $\varkappa$ small
and $\mathrm{mes}\left\{  K\right\}  \geq\rho/2.$ As we show below (see
$\left(  \ref{2.1}\right)  $), the choice of the parameters $R,\rho,U,u$ and
$\varkappa$ is possible, which guarantees the striction transition to happen.

The Hamiltonian (\ref{04}) has the Reflection Positivity (RP) property with
respect to reflections in the shifted coordinate planes:
\[
L_{i;k}=\left\{  x\in\mathbb{R}^{d}:x_{i}=k\right\}  ,\;i=1,...,d,
\]
with integer $k$; it is also RP with respect to reflections in the diagonal
planes
\[
L_{i,j;k}=\left\{  x\in\mathbb{R}^{d}:x_{i}-x_{j}=k\right\}
,\;i,j=1,...,d,\,i\neq j,
\]
again for $k$ integer. To simplify the computations we will use the latter;
this, however, is applicable only in 2D case. The general case can also be
treated, using the RP in coordinate planes, along the same lines.

To formulate our results, we introduce the indicators of some events:

for a bond $l=st$ we define
\[
P_{l}^{<}\left(  \mathbf{r,\sigma}\right)  =\left\{
\begin{array}
[c]{ll}%
1 & \text{ if }r_{l=st}\leq\rho,\\
0 & \text{ otherwise,}%
\end{array}
\right.
\]
where $\rho>0$ is a parameter to be chosen later. Similarly, we define the
indicator
\[
P_{l}^{>}\left(  \mathbf{r,\sigma}\right)  =\left\{
\begin{array}
[c]{ll}%
1 & \text{ if }r_{l=st}\geq\rho+\varepsilon,\\
0 & \text{ otherwise,}%
\end{array}
\right.
\]
where $\varepsilon>0$ is another parameter to be chosen.

We call a Gibbs state $\left\langle \cdot\right\rangle _{\beta},$
corresponding to the Hamiltonian $\left(  \ref{04}\right)  $ and inverse
temperature $\beta,$ a \textbf{contracted} state, iff for every $l$
\[
\left\langle P_{l}^{<}\right\rangle _{\beta}\geq\frac{3}{4}.
\]
Likewise, we call a Gibbs state $\left\langle \cdot\right\rangle _{\beta}$ an
\textbf{expanded} state, iff for every $l$
\[
\left\langle P_{l}^{>}\right\rangle _{\beta}\geq\frac{3}{4}.
\]

\begin{theorem}
Let $h=0.$ It is possible to choose the parameters of the Hamiltonian $\left(
\ref{04}\right)  $ in such a way, that the following holds.

\begin{itemize}
\item at all temperatures low enough there exists a contracted Gibbs state;

\item at all temperatures high enough there exists an expanded Gibbs state;

\item for some critical temperature $\beta_{c}$ there exist at least two
different Gibbs states, $\left\langle \cdot\right\rangle _{\beta_{c}}^{cn}$
and $\left\langle \cdot\right\rangle _{\beta_{c}}^{ex};$ the state
$\left\langle \cdot\right\rangle _{\beta_{c}}^{cn}$ is contracted, while the
state $\left\langle \cdot\right\rangle _{\beta_{c}}^{ex}$ is expanded;

\item if there exists a contracted state at the temperature $\beta^{-1},$ then
in fact there are at least two such states, $\left\langle \cdot\right\rangle
_{\beta}^{+}$ and $\left\langle \cdot\right\rangle _{\beta}^{-}.$ They are
oppositely magnetized: for every $s,t$%
\[
\left\langle \sigma_{s}\right\rangle _{\beta}^{+}=-\left\langle \sigma
_{t}\right\rangle _{\beta}^{-}\geq\frac{3}{4}.
\]

\end{itemize}
\end{theorem}

Our result makes the following \textit{conjectures} very plausible:

\begin{itemize}
\item above the critical temperature $T_{c}$\ every Gibbs state of our
Hamiltonian is expanded, having zero magnetization,

\item below $T_{c}$\ every Gibbs state is contracted, while every pure state
has non-zero magnetization,

\item at $T=T_{c}$\ precisely three pure states coexist: one is expanded, with
zero magnetization, while the other two are contracted and oppositely magnetized.
\end{itemize}

\section{Basic estimates and proof of the main result\label{section 2}}

Our strategy of the proof is to follow the RP theory of the first-order phase
transitions. To this end we introduce the following indicators:
\[%
\begin{array}
[c]{lll}%
P_{l}^{<} & \text{ - of the event } & \{r_{l=st}\leq\rho\},\\
P_{l}^{0} & \text{ - of the event } & \{\rho<r_{l=st}<\rho+\varepsilon\},\\
P_{l}^{>} & \text{ - of the event } & \{r_{l=st}\geq\rho+\varepsilon\},
\end{array}
\]
so $P_{l}^{<}+P_{l}^{0}+P_{l}^{>}=1$. We also introduce the indicators
$P_{\Lambda}^{<},P_{\Lambda}^{0}$ and $P_{\Lambda}^{>},$ which are products of
the above over all bonds, i.e. $P_{\Lambda}^{<}=\prod_{l\in\Lambda_{b}}%
P_{l}^{<},$ etc. We put $P_{\Lambda}^{0>}=\prod_{l\in\Lambda_{b}}\left(
P_{l}^{0}+P_{l}^{>}\right)  .$

The strategy consists in showing that for the finite volume states
$\left\langle \cdot\right\rangle _{\beta}$ with periodic boundary conditions
at inverse temperature $\beta,$ uniformly in volume:

\begin{itemize}
\item the expectation $\left\langle P_{l}^{>}\right\rangle _{\beta}$ is small
at low temperatures,

\item the expectation $\left\langle P_{l}^{<}\right\rangle _{\beta}$ is small
at high temperatures,

\item the expectation $\left\langle P_{l}^{<}P_{l^{\prime}}^{>}\right\rangle
_{\beta}$ is small at all temperatures and for all pairs of bonds $l\neq
l^{\prime},$

\item the expectation $\left\langle P_{l}^{0}\right\rangle _{\beta}$ is small
at all temperatures.
\end{itemize}

\noindent The rest then is standard, see \cite{S}.

(1) First we show that the expectation $\left\langle P_{l}^{>}\right\rangle
_{\beta}$ is small at low temperatures.
\[
\left\langle P_{l}^{>}\right\rangle _{\beta}\leq\left\langle P_{\Lambda}%
^{>}\right\rangle _{\beta}^{1/2\left|  \Lambda\right|  }\leq\left\{
\frac{\left\langle P_{\Lambda}^{>}\right\rangle _{\beta}}{\left\langle
P_{\Lambda}^{<}\right\rangle _{\beta}}\right\}  ^{1/2\left|  \Lambda\right|
}.
\]
We have :
\begin{align*}
\left\langle P_{\Lambda}^{>}\right\rangle _{\beta}  &  =\frac{1}{Z_{\Lambda
}(\beta)}\sum_{\sigma^{\Lambda}}\int\prod_{l\in\Lambda_{b}}dr_{l}I_{\left\{
r_{l}\geq\rho+\varepsilon\right\}  }(r_{l})e^{-\beta H_{\Lambda}%
(\sigma^{\Lambda},r^{\Lambda_{b}})}\\
&  \leq\frac{1}{Z_{\Lambda}(\beta)}\sum_{\sigma^{\Lambda}}\prod_{l\in
\Lambda_{b}}\int_{\rho+\varepsilon}^{\infty}dr_{l}e^{-\beta\mu(r_{l}%
-R)^{2}+\beta u}\leq\frac{1}{Z_{\Lambda}(\beta)}2^{\left|  \Lambda\right|
}\left(  \sqrt{\frac{\pi}{\beta\mu}}\right)  ^{2\left|  \Lambda\right|
}e^{2\left|  \Lambda\right|  \beta u}.
\end{align*}
Here and in the following we use the identity: $\int_{-\infty}^{+\infty
}e^{-Ax^{2}}\,dx=\sqrt{\frac{\pi}{A}}.$ On the other hand
\begin{align}
\left\langle P_{\Lambda}^{<}\right\rangle _{\beta}  &  =\frac{1}{Z_{\Lambda
}(\beta)}\sum_{\sigma^{\Lambda}}\int\prod_{l\in\Lambda_{b}}dr_{l}I_{\left\{
r_{l}\leq\rho\right\}  }(r_{l})e^{-\beta H_{\Lambda}(\sigma^{\Lambda
},r^{\Lambda_{b}})}\label{05}\\
&  \geq\frac{1}{Z_{\Lambda}(\beta)}e^{2\beta U\left|  \Lambda\right|  }%
\prod_{l\in\Lambda_{b}}\int_{K}dr_{l}e^{-\beta\mu R^{2}}e^{-2\beta\lambda
\rho^{2}}\nonumber\\
&  =\frac{1}{Z_{\Lambda}(\beta)}\left(  \frac{\rho}{2}e^{\beta(U-\mu
R^{2}-2\lambda\rho^{2})}\right)  ^{2\left|  \Lambda\right|  }.\nonumber
\end{align}
Therefore,
\begin{equation}
\left\langle P_{l}^{>}\right\rangle _{\beta}\leq\frac{2\sqrt{2\pi}%
e^{-\beta(U-u-\mu R^{2}-2\lambda\rho^{2})}}{\rho\sqrt{\beta\mu}}, \label{2.11}%
\end{equation}
which is small for $\beta$ large once
\begin{equation}
U-u>\mu R^{2}+2\lambda\rho^{2}. \label{2.12}%
\end{equation}

(2) Next we show that the expectation $\left\langle P_{l}^{<}\right\rangle
_{\beta}$ is small at high temperatures:
\[
\left\langle P_{l}^{<}\right\rangle _{\beta}\leq\left\langle P_{\Lambda}%
^{<}\right\rangle _{\beta}^{1/2\left\vert \Lambda\right\vert }\leq\left\{
\frac{\left\langle P_{\Lambda}^{<}\right\rangle _{\beta}}{\left\langle
P_{\Lambda}^{0>}\right\rangle _{\beta}}\right\}  ^{1/2\left\vert
\Lambda\right\vert }.
\]
We have:
\begin{align*}
\left\langle P_{\Lambda}^{<}\right\rangle _{\beta}  &  =\frac{1}{Z_{\Lambda
}(\beta)}\sum_{\sigma^{\Lambda}}\int\prod_{l\in\Lambda_{b}}dr_{l}I_{\left\{
r_{l}\leq\rho\right\}  }(r_{l})e^{-\beta H_{\Lambda}(\sigma^{\Lambda
},r^{\Lambda_{b}})}\\
&  \leq\frac{1}{Z_{\Lambda}(\beta)}\left(  \sqrt{2}\rho e^{\beta\left(
\bar{U}-\mu(\rho-R)^{2}\right)  }\right)  ^{2\left\vert \Lambda\right\vert }.
\end{align*}
For the lower bound we have
\begin{align}
\left\langle P_{\Lambda}^{0>}\right\rangle _{\beta}  &  =\frac{1}{Z_{\Lambda
}(\beta)}\sum_{\sigma^{\Lambda}}\int\prod_{l\in\Lambda_{b}}dr_{l}I_{\left\{
r_{l}\geq\rho\right\}  }(r_{l})e^{-\beta H_{\Lambda}(\sigma^{\Lambda
},r^{\Lambda_{b}})}\label{06}\\
&  \geq\frac{1}{Z_{\Lambda}(\beta)}2^{\left\vert \Lambda\right\vert }\int
\prod_{l\in\Lambda_{b}}dr_{l}I_{\left\{  r_{l}\geq\rho\right\}  }%
(r_{l})e^{-\beta\left(  u+\mu\left(  r_{l}-R\right)  ^{2}+\lambda
\sum_{l^{\prime}:l\text{ nn }l^{\prime}}\left(  r_{l}-r_{l^{\prime}}\right)
^{2}\right)  }\nonumber\\
&  \geq\frac{1}{Z_{\Lambda}(\beta)}2^{\left\vert \Lambda\right\vert }%
\prod_{l\in\Lambda_{b}}\int_{-\left(  R-\rho\right)  }^{\infty}dr_{l}%
e^{-\beta\left(  \mu+8\lambda\right)  r_{l}{}^{2}-\beta u}\nonumber\\
&  \geq\frac{1}{Z_{\Lambda}(\beta)}\left(  \sqrt{\frac{\pi}{\beta\left(
\mu+8\lambda\right)  }}e^{-\beta u}\right)  ^{2\left\vert \Lambda\right\vert
},\nonumber
\end{align}
where we use in the third line the inequality $\left(  x-y\right)  ^{2}%
\leq2x^{2}+2y^{2},$ and also the fact that for every $l$ the sum
$\sum_{l^{\prime}:l\text{ nn }l^{\prime}}\left(  r_{l}-r_{l^{\prime}}\right)
^{2}$ has $4$ terms. Therefore
\begin{equation}
\left\langle P_{l}^{<}\right\rangle _{\beta}\leq\sqrt{\frac{2\beta\left(
\mu+8\lambda\right)  }{\pi}}\rho e^{\beta\left(  \bar{U}+u-\mu(\rho
-R)^{2}\right)  }, \label{2.21}%
\end{equation}
which is small for small $\beta.$

(3) Now we estimate the correlation function $\left\langle P_{l}%
^{<}P_{l^{\prime}}^{>}\right\rangle _{\beta},$ with $l=\langle st\rangle,$
$l^{\prime}=\langle s^{\prime}t\rangle,$ $\left|  s-s^{\prime}\right|
=\sqrt{2}.$ We have:
\[
\left\langle P_{l}^{<}P_{l^{\prime}}^{>}\right\rangle _{\beta}\leq\left\langle
P_{\Lambda}^{\gtrless}\right\rangle _{\beta}^{1/\left|  \Lambda\right|  },
\]
where the indicator $P_{\Lambda}^{\gtrless}$ corresponds to the following
event: on half of the bonds -- $\Lambda_{b}^{\geq}$ -- of $\Lambda_{b}$ -
namely, on those which have one endpoint on the sublattice, generated by the
vectors $\left(  1,1\right)  $ and $\left(  2,-2\right)  $ - the event
$r_{\cdot}\geq\rho+\varepsilon$ happens, while on the remaining ones --
$\Lambda_{b}^{\leq}=\Lambda_{b}\,\backslash\,\Lambda_{b}^{\geq}$ -- the event
$r_{\cdot}\leq\rho$ happens. Therefore
\begin{align*}
\left\langle P_{\Lambda}^{\gtrless}\right\rangle _{\beta}  &  =\frac
{1}{Z_{\Lambda}(\beta)}\sum_{\sigma^{\Lambda}}\int e^{-\beta H_{\Lambda
}(\sigma^{\Lambda},r^{\Lambda_{b}})}\prod_{l\in\Lambda_{b}^{\geq}}%
\prod_{l^{\prime}\in\Lambda_{b}^{\leq}}dr_{l}I_{\left\{  r_{l}\geq
\rho+\varepsilon\right\}  }(r_{l})dr_{l^{\prime}}I_{\left\{  r_{l^{\prime}%
}\leq\rho\right\}  }(r_{l^{\prime}})\\
&  \leq\frac{1}{Z_{\Lambda}(\beta)}2^{\left|  \Lambda\right|  }e^{\beta\left(
\bar{U}+u-\mu(\rho-R)^{2}-\lambda\varepsilon^{2}\right)  \left|
\Lambda\right|  }\rho^{\left|  \Lambda\right|  }\left(  \int_{\rho
+\varepsilon}^{\infty}dre^{-\beta\mu(r-R)^{2}}\right)  ^{\left|
\Lambda\right|  }\\
&  \leq\frac{1}{Z_{\Lambda}(\beta)}\left(  2\sqrt{\frac{\pi}{\beta\mu}}\rho
e^{\beta\left(  \bar{U}+u-\mu(\rho-R)^{2}-\lambda\varepsilon^{2}\right)
}\right)  ^{\left|  \Lambda\right|  }.
\end{align*}

To estimate the partition function from below we note that $P_{\Lambda}%
^{<}\left(  \cdot\right)  +P_{\Lambda}^{0>}\left(  \cdot\right)  \leq1,$ so
\begin{equation}
Z_{\Lambda}(\beta)\geq Z_{\Lambda}(\beta)\left(  \left\langle P_{\Lambda}%
^{<}\right\rangle _{\beta}+\left\langle P_{\Lambda}^{0>}\right\rangle _{\beta
}\right)  . \label{03}%
\end{equation}
Using (\ref{05}), (\ref{06}), we thus have
\begin{equation}
\left\langle P_{\Lambda}^{\gtrless}\right\rangle _{\beta}\leq\frac{\left(
2\sqrt{\frac{\pi}{\beta\mu}}\rho e^{\beta\left(  \bar{U}+u-\mu(\rho
-R)^{2}-\lambda\varepsilon^{2}\right)  }\right)  ^{\left\vert \Lambda
\right\vert }}{\left(  \frac{\rho}{2}e^{\beta(U-\mu R^{2}-2\lambda\rho^{2}%
)}\right)  ^{2\left\vert \Lambda\right\vert }+\left(  \sqrt{\frac{\pi}%
{\beta\left(  \mu+8\lambda\right)  }}e^{-\beta u}\right)  ^{2\left\vert
\Lambda\right\vert }}. \label{2.41}%
\end{equation}

By suppressing one of the terms in the denominator of (\ref{2.41}) we get the
following two estimates:
\begin{equation}
\left\langle P_{l}^{<}P_{l^{\prime}}^{>}\right\rangle _{\beta}\leq\frac
{8}{\rho}\sqrt{\frac{\pi}{\beta\mu}}e^{\beta\left(  \bar{U}+u-2U-\mu
(\rho-R)^{2}+2\mu R^{2}-\lambda\varepsilon^{2}+4\lambda\rho^{2}\right)  },
\label{2.43}%
\end{equation}
which is good for $\beta$ large, and
\begin{equation}
\left\langle P_{l}^{<}P_{l^{\prime}}^{>}\right\rangle _{\beta}\leq2\sqrt
{\frac{\beta}{\pi\mu}}\rho\left(  \mu+8\lambda\right)  e^{\beta\left(  \bar
{U}+3u-\mu(\rho-R)^{2}-\lambda\varepsilon^{2}\right)  }, \label{2.45}%
\end{equation}
which is good for $\beta$ small. So we have to look for some intermediate
value of $\beta^{\ast},$ such that for $\beta\geq\beta^{\ast}$ the r.h.s. of
(\ref{2.43}) is small, while for $\beta\leq\beta^{\ast}$ the r.h.s. of
(\ref{2.45}) is small. Of course, such value of the inverse temperature should
be the one which makes the two terms in the denominator of (\ref{2.41}) equal;
in other words, the reasonable choice of the value $\beta^{\ast}$ is to take
it to be the solution of the equation
\begin{equation}
\frac{\rho}{2}e^{\beta(U-\mu R^{2}-2\lambda\rho^{2})}=\sqrt{\frac{\pi}%
{\beta\left(  \mu+8\lambda\right)  }}e^{-\beta u}. \label{20}%
\end{equation}
But any choice of $\beta^{\ast}$ would be as good as this one, provided only
that the estimates (\ref{2.43}) and (\ref{2.45}) will turn into bounds strong enough.

(4) The last estimate we need is that for the expectation $\left\langle
P_{l}^{0}\right\rangle _{\beta}.$ We have
\[
\left\langle P_{l}^{0}\right\rangle _{\beta}\leq\left\langle P_{\Lambda}%
^{0}\right\rangle _{\beta}^{1/2\left\vert \Lambda\right\vert }.
\]
Now
\begin{align*}
\left\langle P_{\Lambda}^{0}\right\rangle _{\beta}  &  =\frac{1}{Z_{\Lambda
}(\beta)}\sum_{\sigma^{\Lambda}}\int\prod_{l\in\Lambda_{b}}dr_{l}I_{\left\{
\rho<r_{l}<\rho+\varepsilon\right\}  }(r_{l})e^{-\beta H_{\Lambda}%
(\sigma^{\Lambda},r^{\Lambda_{b}})}\\
&  \leq\frac{1}{Z_{\Lambda}(\beta)}2^{\left\vert \Lambda\right\vert }%
(\int_{\rho}^{\rho+\varepsilon}dre^{-\beta\left[  \mu(\rho+\varepsilon
-R)^{2}-u\right]  })^{2\left\vert \Lambda\right\vert }\\
&  \leq\frac{1}{Z_{\Lambda}(\beta)}\left(  \sqrt{2}\varepsilon e^{-\beta
\left[  \mu(\rho+\varepsilon-R)^{2}-u\right]  }\right)  ^{2\left\vert
\Lambda\right\vert }.
\end{align*}
Combining with the estimate (\ref{03}) we find:
\begin{equation}
\left\langle P_{\Lambda}^{0}\right\rangle _{\beta}\leq\frac{\left(  \sqrt
{2}\varepsilon e^{-\beta\left[  \mu(\rho+\varepsilon-R)^{2}-u\right]
}\right)  ^{2\left\vert \Lambda\right\vert }}{\left(  \frac{\rho}{2}%
e^{\beta(U-\mu R^{2}-2\lambda\rho^{2})}\right)  ^{2\left\vert \Lambda
\right\vert }+\left(  \sqrt{\frac{\pi}{\beta\left(  \mu+8\lambda\right)  }%
}e^{-\beta u}\right)  ^{2\left\vert \Lambda\right\vert }}. \label{01}%
\end{equation}
Here we can proceed as in the previous case, turning (\ref{01}) into two
different estimates, depending on the value of $\beta.$ However, the case of
the observable $P_{l}^{0}$ is easier, and it is sufficient to keep just one
summand in the denominator of (\ref{01}) in order to get a reasonable estimate
on it. Namely, we keep the second one, arriving to
\begin{equation}
\left\langle P_{l}^{0}\right\rangle _{\beta}\leq\sqrt{\frac{2\beta\left(
\mu+8\lambda\right)  }{\pi}}\varepsilon e^{-\beta\left[  \mu(\rho
+\varepsilon-R)^{2}-2u\right]  }. \label{2.35}%
\end{equation}

We now shall show that if we make for the Hamiltonian (\ref{04}) the following
choice of the interaction parameters :
\begin{equation}
\lambda=\mu=1,U=2R^{2},\bar{U}=\left(  2+\delta^{2}\right)  R^{2}%
,u=\delta,\rho=R^{-1},\varepsilon=2\delta R, \label{2.1}%
\end{equation}
with $R$ big enough and $\delta$ small enough, then the conditions of our
theorem hold for the interval $\left[  \beta_{+},\beta_{-}\right]  ,$ provided
$\beta_{+}=\beta_{+}\left(  R,\delta\right)  $ is small enough, and $\beta
_{-}=\beta_{-}\left(  R,\delta\right)  $ is large enough.

Since the estimate (\ref{2.12}) is satisfied under our choice (\ref{2.1}), the
relation (\ref{2.11}) holds for all $\beta$ large enough. As we said before,
the r.h.s. of $\left(  \ref{2.21}\right)  $ is small for all $\beta$ small
enough. Therefore it is enough to check that the r.h.s. of (\ref{2.41}) and
(\ref{01}) are small uniformly in all $\beta.$

To proceed with the estimate of the correlation function $\left\langle
P_{l}^{<}P_{l^{\prime}}^{>}\right\rangle _{\beta},$ as indicated above, we
have to choose a value of the intermediate inverse temperature $\beta^{\ast}.$
Our choice is%

\begin{equation}
\beta^{\ast}\mathbf{\ }=\mathbf{\ }2\,R^{-2}\ln R. \label{10}%
\end{equation}
\textbf{\ \ }One can check that thus defined\textbf{\ }$\beta^{\ast}$ is
indeed an approximate solution to (\ref{20}) \textbf{as }$R\rightarrow\infty,$
though this is not important.\textbf{\ \ }

In the region\textbf{\ }$\beta\geq\beta^{\ast}$ we will use the estimate
(\ref{2.43}), which under the choice (\ref{2.1}) becomes
\begin{align*}
\left\langle P_{l}^{<}P_{l^{\prime}}^{>}\right\rangle _{\beta}  &  \leq
8R\sqrt{\frac{\pi}{\beta^{\ast}}}e^{\beta^{\ast}\left(  \left(  2+\delta
^{2}\right)  R^{2}+\delta-4R^{2}-(R-R^{-1})^{2}+2R^{2}-4\delta^{2}%
R^{2}+4R^{-2}\right)  }\\
&  \leq8R\sqrt{\frac{\pi}{\beta^{\ast}}}e^{\beta^{\ast}\left(  -\left(
1+2\delta^{2}\right)  R^{2}\right)  }\\
&  \leq8R^{2}\sqrt{\frac{\pi}{2\,\ln R}}R^{-2\,\left(  1+2\delta^{2}\right)
}\\
&  \leq R^{-4\delta^{2}}%
\end{align*}
for $R$ large.

In the region\textbf{\ }$\beta\leq\beta^{\ast}$ we shall use the estimate
(\ref{2.45}), which similarly becomes
\begin{align*}
\left\langle P_{l}^{<}P_{l^{\prime}}^{>}\right\rangle _{\beta}  &  \leq
18\sqrt{\beta\pi^{-1}}R^{-1}e^{\beta\left(  \left(  2+\delta^{2}\right)
R^{2}+3\delta-(R-R^{-1})^{2}-4\delta^{2}R^{2}\right)  }\\
&  \leq18\sqrt{\beta\pi^{-1}}R^{-1}e^{\beta\left(  1-2\delta^{2}\right)
R^{2}}\\
&  \leq18\sqrt{\beta^{\ast}\pi^{-1}}R^{-1}e^{\beta^{\ast}\left(  1-2\delta
^{2}\right)  R^{2}}\\
&  \leq18\sqrt{2\pi^{-1}\mathbf{\,}\ln R}R^{-4\delta^{2}}\\
&  \leq R^{-3\delta^{2}}%
\end{align*}
for $R$ large.

Finally we consider the bound (\ref{2.35}), which becomes
\begin{align*}
\left\langle P_{l}^{0}\right\rangle _{\beta}  &  \leq6\sqrt{\frac{2\beta}{\pi
}}\delta Re^{-\beta\left[  (R^{-1}+2\delta R-R)^{2}-2\delta\right]  }\\
&  \leq6\sqrt{\frac{2\beta}{\pi}}\delta Re^{-\beta(1-3\delta)^{2}R^{2}%
}\mathbf{.}%
\end{align*}
Note that the function $\sqrt{x}e^{-ax}$ has its maximum at $x=\frac{1}{2a},$
which equals to $\sqrt{\frac{1}{2ea}}.$ Applying this to the last expression
with $x=\beta R^{2},$ we get
\[
\left\langle P_{l}^{0}\right\rangle _{\beta}\leq6\frac{\delta}{1-3\delta}%
\sqrt{\frac{1}{\pi e}},
\]
which is small for small $\delta$ at any $\beta.$

\section{Magnetization}

Here we will prove the last statement of our theorem: the occurrence of
spontaneous magnetization in the contracted states. To do this we split the
event $\{r_{l=st}\leq\rho\}$ into four events, and we introduce the
corresponding four indicators
\[%
\begin{array}
[c]{lll}%
P_{l}^{<\pm\pm} & \text{ - of the event } & \{r_{l=st}\leq\rho,\sigma_{s}%
=\pm,\sigma_{t}=\pm\}.
\end{array}
\]

We will show now that for all $\beta$ the expectations $\left\langle
P_{l}^{<+-}\right\rangle _{\beta}=\left\langle P_{l}^{<-+}\right\rangle
_{\beta}$ are small, uniformly in the volume. Together with the obvious
statements that
\[
\left\langle P_{l}^{<++}\right\rangle _{\beta}=\left\langle P_{l}%
^{<--}\right\rangle _{\beta}%
\]
and
\[
\left\langle P_{l}^{<++}\right\rangle _{\beta}+\left\langle P_{l}%
^{<--}\right\rangle _{\beta}+\left\langle P_{l}^{<+-}\right\rangle _{\beta
}+\left\langle P_{l}^{<-+}\right\rangle _{\beta}=\left\langle P_{l}%
^{<}\right\rangle _{\beta},
\]
that implies our claim, due to the first part of our theorem and by subsequent
application of the Theorem XX of \cite{S}.

We have
\[
\left\langle P_{l}^{<-+}\right\rangle _{\beta}\leq\left\langle P_{\Lambda
}^{<-+}\right\rangle _{\beta}^{1/2\left\vert \Lambda\right\vert },
\]
where $P_{\Lambda}^{<-+}$ the indicator of the event that for every bond
$l^{\prime}$ we have $r_{l^{\prime}}\leq\rho,$ while
\[
\sigma_{s}=\left\{
\begin{array}
[c]{ll}%
+1 & \text{ if }\left\vert s_{1}\right\vert +\left\vert s_{2}\right\vert
\text{ is even,}\\
-1 & \text{ otherwise.}%
\end{array}
\right.
\]
We denote this spin arrangement by $\sigma_{\pm}^{\Lambda}.$ So
\begin{align*}
\left\langle P_{\Lambda}^{<-+}\right\rangle _{\beta}  &  =\frac{1}{Z_{\Lambda
}(\beta)}\int e^{-\beta H_{\Lambda}(\sigma_{\pm}^{\Lambda},r^{\Lambda_{b}}%
)}\prod_{l\in\Lambda_{b}}dr_{l}I_{\left\{  r_{l}\leq\rho\right\}  }(r_{l})\\
&  \leq\frac{1}{Z_{\Lambda}(\beta)}\left(  \rho e^{-\beta\mu(\rho-R)^{2}%
}\right)  ^{2\left\vert \Lambda\right\vert }.
\end{align*}
As in $\left(  \ref{2.41}-\ref{2.45}\right)  ,$ we have two estimates:
\begin{equation}
\left\langle P_{l}^{<-+}\right\rangle _{\beta}\leq\frac{\rho e^{-\beta\mu
(\rho-R)^{2}}}{\frac{\rho}{2}e^{\beta(U-\mu R^{2}-2\lambda\rho^{2})}%
}=2e^{-\beta\left[  U+\mu(\rho-R)^{2}-\mu R^{2}-2\lambda\rho^{2}\right]  }
\label{31}%
\end{equation}
and
\begin{equation}
\left\langle P_{l}^{<-+}\right\rangle _{\beta}\leq\frac{\rho e^{-\beta\mu
(\rho-R)^{2}}}{\sqrt{\frac{\pi}{\beta\left(  \mu+8\lambda\right)  }}e^{-\beta
u}}=\rho\sqrt{\frac{\beta\left(  \mu+8\lambda\right)  }{\pi}}e^{-\beta\left[
\mu(\rho-R)^{2}-u\right]  }. \label{32}%
\end{equation}
In fact, with our choice $\left(  \ref{2.1}\right)  $ of the parameters the
second one is effective for all $\beta.$ We have
\[
\left\langle P_{l}^{<-+}\right\rangle _{\beta}\leq3R^{-1}\sqrt{\frac{\beta
}{\pi}}e^{-\beta\left[  (R-R^{-1})^{2}-\delta\right]  }.
\]
The r.h.s. has its maximum at $\beta=\frac{1}{2\left[  (R-R^{-1})^{2}%
-\delta\right]  },$ so
\[
\left\langle P_{l}^{<-+}\right\rangle _{\beta}\leq3R^{-1}\sqrt{\frac{1}{2\pi
e\left[  (R-R^{-1})^{2}-\delta\right]  }},
\]
which is small for $R$ large enough.$\blacksquare$

\end{document}